\documentclass[prd,twocolumn,superscriptaddress]{revtex4-1}
\usepackage{epsfig}
\usepackage{graphicx}
\usepackage{palatino}
\usepackage[english]{babel}
\usepackage{hyphenat}
\usepackage{amsmath}
\usepackage{amssymb}
\usepackage{mathtools}
\usepackage{mathrsfs}
\usepackage{slashed}
\usepackage{epstopdf}
\usepackage{xcolor}
\usepackage{booktabs}
\definecolor{lcolor}{rgb}{0.,0.0,0.}
\definecolor{citcolor}{rgb}{0,0.,0.5}
\usepackage[breaklinks,colorlinks,urlcolor=blue,citecolor=blue,linkcolor=blue]{hyperref}
\usepackage{multirow}
\usepackage{ltablex}
\usepackage{soul}

\newcommand{\beq}{\begin{eqnarray}}
\newcommand{\eeq}{\end{eqnarray}}
\newcommand{\dis}{\displaystyle}

\newcommand{\bem}{\begin{multline}}
\newcommand{\eem}{\end{multline}}
\newcommand{\beg}{\begin{gather}}
\newcommand{\eeg}{\end{gather}}

\newcommand{\nn}{\nonumber\\}

\newcommand{\ben}{\begin{eqnarray*}}
\newcommand{\een}{\end{eqnarray*}}

\newcommand{\eqn}[1]{Eq.~\eqref{#1}}

\def\cP{{\cal P}}

\def\cO{{\cal O}}

\newcommand{\secn}[1]{Section~1}
\newcommand{\appn}[1]{Appendix~1}

\long\def\comment#1{ }

\def\and{\quad\text{and}\quad}

\newcommand{\rmd}{{\rm d}}

\newcommand{\rme}{{\rm e}}

\def\q{{\boldsymbol q}}
\def\0{{\boldsymbol 0}}

\def\k{{\boldsymbol k}}

\def\x{{\boldsymbol x}}

\newcommand{\del}{\partial}

\DeclareMathOperator*{\sumint}{%
\mathchoice%
  {\ooalign{$\displaystyle\sum$\cr\hidewidth$\displaystyle\int$\hidewidth\cr}}
  {\ooalign{\raisebox{.14\height}{\scalebox{.7}{$\textstyle\sum$}}\cr\hidewidth$\textstyle\int$\hidewidth\cr}}
  {\ooalign{\raisebox{.2\height}{\scalebox{.6}{$\scriptstyle\sum$}}\cr$\scriptstyle\int$\cr}}
  {\ooalign{\raisebox{.2\height}{\scalebox{.6}{$\scriptstyle\sum$}}\cr$\scriptstyle\int$\cr}}
}

\begin{document}

\title{Revisiting transverse momentum broadening in dense QCD media}

\author{Jo\~ao Barata}
\email[]{joaolourenco.henriques@usc.es}
\affiliation{Instituto Galego de Fisica de Altas Enerxias (IGFAE), Universidade de Santiago de Compostela,E-15782 Galicia, Spain}
\author{Yacine Mehtar-Tani}
\email[]{mehtartani@bnl.gov}
\affiliation{Physics Department, Brookhaven National Laboratory, Upton, NY 11973, USA}
\affiliation{RIKEN BNL Research Center, Brookhaven National Laboratory, Upton, NY 11973, USA}
\author{Alba Soto-Ontoso}
\email[]{ontoso@bnl.gov}
\affiliation{Physics Department, Brookhaven National Laboratory, Upton, NY 11973, USA}
\author{Konrad Tywoniuk}
\email[]{konrad.tywoniuk@uib.no}
\affiliation{Department of Physics and Technology, University of Bergen, 5007 Bergen, Norway}

\begin{abstract}
We reconsider the problem of transverse momentum broadening of a highly-energetic parton suffering multiple scatterings in dense colored media, such as the thermal Quark-Gluon plasma or large nuclei. In the framework of Moli\`ere's theory of multiple scattering we re-derive a simple analytic formula, to be used in jet quenching phenomenology, that accounts for both the multiple soft and hard Rutherford scattering regimes. Further, we discuss the sensitivity of momentum broadening to modeling of the non-perturbative infrared sector by presenting a detailed analytic and numerical comparison between the two widely used models in phenomenology: the Hard Thermal Loop and the Gyulassy-Wang potentials. We show that for the relevant values of the parameters the non-universal, model dependent contributions are negligible at LHC, RHIC and EIC energies, thus consolidating the predictive power of jet quenching theory. 
\end{abstract}

\maketitle

\section{Introduction}
\label{sec:intro}
The propagation of highly energetic partons in a dense QCD medium is affected by multiple collisions with medium constituents that cause their transverse momentum distribution to broaden. In addition, medium-induced gluon emissions can be triggered by these random kicks leading to the formation of a parton cascade.

The latter is the main mechanism of parton energy loss in large QCD media \cite{Blaizot_review,Kurkela} and its theoretical description has drawn a lot of attention in the last two decades.
However, most analytic results used in phenomenology were obtained in certain kinematic limits~\cite{BDMPS1,BDMPS2,BDMPS3,BDMPS4,Wiedemann,GLV} and a solid and quantitative understanding of in-medium jet modifications, as measured in experiment, requires further theoretical control over the full kinematic range of the transverse momentum dynamics. This is currently of particular importance as the field moves towards highly precise comparisons between data and theoretical predictions.

In the present manuscript we study the single particle momentum broadening distribution for the case of an energetic parton propagating through a homogeneous plasma brick of length $L$. A complementary study of the medium induced radiation mechanism, following a similar scheme to the one present in this paper, is discussed in \cite{Blok:2020jgo,BMST} for the transverse momentum dependent gluon spectrum and in \cite{IOE1,IOE2,IOE3} for its integrated version.

Although substantial effort has been put into studying momentum broadening in increasingly more realistic and complex scenarios \cite{Ipp,Romatschke,Baier,Moore:2019lgw,vanHameren:2019xtr,DEramo:2012uzl,DEramo:2010wup,Broad_EFT,Gyulassy:2002yv,Qiu:2003pm,Dumitru:2007rp}, theoretical uncertainties in the simplistic setup considered here remain to be understood. The origin of these ambiguities is mainly associated to the infrared (IR) modeling of the parton-medium interaction. In addition, it is common in phenomenological studies to take the broadening distribution in some limiting form such as the Gaussian approximation that applies to the multiple soft scattering regime. Even though the latter does not capture the correct physics in the whole range of transverse momenta, its simple analytical form makes it suitable for theoretical calculations, e.g.~\cite{Ringer:2019rfk,Mueller:2016gko}, together with a straightforward implementation in jet quenching Monte-Carlo event generators, where a given momentum broadening probability has to be sampled~\cite{Armesto:2009fj,Hybrid,Saclay}.  

The goal of this paper is twofold. First, we revisit the theory of multiple scattering by Moli\`ere \cite{Moliere,Bethe:1953va} which provides a more accurate description of the broadening distribution over the full range in transverse momentum. In particular, the Gaussian behavior and the power-law Coulomb tail, pertaining to Rutherford scattering, are reproduced at small and large transverse momentum, respectively. On the other hand, we systematically study the role of non-perturbative modeling on the momentum broadening probability distribution. We show that, for realistic values of the parameters, resulting non-universal power corrections are negligible. 

This paper is structured as follows. Section~\ref{sec:kinetic} introduces the kinetic theory formulation of the momentum broadening probability distribution, along with the two medium models to be explored. Next, Section~\ref{sec:distribution-ioe} reviews Moli\`ere's systematic approach to multiple scattering theory applied to momentum broadening. In Section~\ref{sec:HT-GW} we explore the dependence of the broadening distribution on IR modeling. In addition, we discuss our findings in the parameter space explored by LHC, RHIC and the upcoming EIC in Section~\ref{sec:numerics}. We summarize our results and discuss possible future directions in Section \ref{sec:summary}. Complementary material is presented in Appendices~\ref{appendix:integrals} to \ref{appendix:kinetic_broad}.

\section{Kinetic description of momentum broadening}
\label{sec:kinetic}
The probability for a highly energetic parton traversing a dense QCD medium to acquire a transverse momentum $\k$ from multiple scattering during a time $L$ is denoted by $\cP(\k,L)$. Its time evolution can be formulated in kinetic theory and is given by~\cite{BDIM1,Blaizot:2012fh}
\begin{align}\label{eq:kin-mom}
\frac{\del }{\del L }  \cP(\k, L) = C_R\int_\q \, \gamma(\q) \,\left[ \cP(\k-\q,L) -\cP(\k,L) \right] \, ,
\end{align}
where $\gamma(\q)$ is the collision rate and is related to the in-medium elastic scattering cross-section to be discussed shortly. $C_R$ is the parton color factor in representation $R$, i.e., $C_R=C_F=(N_c^2-1)/(2N_c)$ and $C_R=C_A=N_c$ for a quark and a gluon, respectively. Here and throughout we used the shorthand notation $\int_\q \equiv \int \rmd^2 \q/(2\pi)^2$ for transverse momentum integration, and $\int_\x\equiv \int d^2\x$ for transverse coordinate space integrals.

Physically, \eqn{eq:kin-mom} has a simple interpretation: in an infinitesimal time step $\delta t$, the probability for a parton to end up with momentum $\k$ at $L+\delta t$ equals the probability of starting with momentum $\k-\q$ and acquiring momentum $\q$ during $\delta t$. In addition, one must subtract the probability of already starting with momentum $\k$ and diffusing to some other momentum mode. This structure ensures the unitarity of $\cP$ and, consequently, its probabilistic interpretation. We remind the reader that the previous result is only valid in the strict eikonal limit, where the acquired transverse momentum is much smaller than the parton energy.

Turning to the collision rate $\gamma(\q)$, a closed expression exists in two distinct scenarios. In the case of a medium formed by static scattering centres with Yukawa-type interactions, $\gamma(\q)$ is given by the so-called Gyulassy-Wang (GW) model~\cite{GW}
\beq\label{eq:GW}
\gamma^{\rm  GW}(\q) = \frac{g^4 n}{(\q^2+\mu^2)^2} \, ,
\eeq
where $n$ represents the density of scattering centres and $\mu$ is the (GW) screening mass.
When the QCD medium is in thermal equilibrium, $\gamma(\q)$ can be computed with Hard Thermal Loop theory (HTL) resulting in~\cite{HTL}
\beq\label{eq:HTL}
\gamma^{\rm  HTL}(\q) = \frac{g^2m_D^2T}{\q^2\,(\q^2+m_D^2)} \,,
\eeq
where $g$ is the strong coupling and $m_D$ corresponds to the Debye mass whose temperature ($T$) dependence is given, at leading order in $g$, by $m^2_D(T)\!=\!(1+\frac{n_{\rm{f}}}{6})g^2T^2$, where $n_{\rm f}$ is the number of active light flavours. While the UV behavior is common to both models, i.e., $\q^{-4}$, the main difference is in the way the IR Coulomb divergence is regulated. In the limit $\q \to 0$ we have, $\gamma^{\rm  GW}(\q)\to {\rm const.}$ while $\gamma^{\rm  HTL}(\q)\to \q^{-2}$. The possibility of mapping $\mu$ and $m_D$ together with other differences and similarities between these two models for $\gamma(\q)$, commonly used in phenomenology, will be discussed in Section~\ref{sec:HT-GW}.\par 
Defining $S(\x,L)$ as the Fourier transform of $\cP(\k,L)$, 
\beq 
S(\x,L) = \int_\q \, \cP(\q,L)\, \rme^{i\q\cdot\x}\,,
\eeq
\eqn{eq:kin-mom} becomes local in position space 
\beq\label{eq:kin-pos}
\frac{\del }{\del L }  S(\x, L) = - v(\x)  S(\x, L)\,  ,
\eeq
and can be easily solved to give
\begin{equation}\label{eq:S_full}
S(\x,L)=\rme^{-\int_0^L ds \, v(\x,s)}=\rme^{-v(\x)\, L}    \, ,
\end{equation}
where in the last step we assumed that the medium is a homogeneous brick of length $L$ and defined the so-called \textit{dipole cross-section} \footnote{The dipole we refer to is formed by Wilson lines in amplitude and conjugate amplitude.}
as
\beq\label{eq:v}
 v(\x)  \equiv C_R\int_\q \left(1-\rme^{i\q\cdot \x}\right)\gamma(\q) \, .
 \eeq
Explicit formulas for $v(\x)$ in the GW and HTL models can be obtained using the integrals computed in Appendix~\ref{appendix:dipole}. They read
\begin{align}\label{eq:v_GW_text}
    v^{\rm GW}(\x)&=\frac{\hat{q}_0}{\mu^2}\left(1-\mu |\x|K_1(\mu|\x|)\right) \, ,
\end{align}
and
\begin{align}\label{eq:v_HTL_text}
    v^{\rm HTL}(\x)&
    & = \frac{2\hat{q}_0}{m_D^2}\left(K_0(m_D|\x|)+\log\left(\frac{m_D|\x|}{2}\right)+\gamma_E\right) \, ,
\end{align} 
where 
\beq 
\hat{q}_0=  \begin{cases}   4\pi \alpha_s^2 C_R n \,\quad &{\text{for  GW}}  \\
 \alpha_s C_R m_D^2 T\,\quad &{\text{ for HTL}}
\end{cases} \, ,
\eeq
is the \textit{bare} jet quenching transport parameter and $\alpha_s=g^2/(4\pi)$.

Finally, by Fourier transforming \eqn{eq:S_full} back into momentum space, the probability for a parton to acquire transverse momentum $\k$ reads
\begin{equation}\label{eq:pt-broad-def}
\cP(\k,L)=\int_\x \, (S(\x,L)-S(\infty,L)) \rme^{-i\x\cdot\k} \, ,
\end{equation}
where we subtract the \textit{no-broadening} contribution that would result in a $\delta^{(2)}(\k)$ term that does not contribute at finite $\k$ (see \cite{Feal:2018jbm}). This term only vanishes when $v(\infty)=+\infty$. From \eqn{eq:v_GW_text} and \eqn{eq:v_HTL_text}, we observe that neither GW nor HTL satisfy this condition, but rather saturate at large $|\x|$. Regarding the normalization of the probability distribution given by \eqn{eq:pt-broad-def}, it is easy to verify, using $v(\textbf{0})\!=\!0$, that $\int_\k \cP(\k,L)=1-\exp(-v(\infty)L)$.\par 

To the best of our knowledge, a closed analytic form for \eqn{eq:pt-broad-def} when using either GW or HTL expressions for $v(\x)$ does not exist. It is the goal of the next Section to achieve an analytic representation of the transverse momentum probability distribution, $\cP(\k,L)$, by combining an 
expansion of \eqn{eq:pt-broad-def} in universal and model dependent terms together with Moli\`ere's theory~\cite{Moliere,Bethe:1953va} of multiple scattering.

\section{$\mathcal P(\k)$-distribution from Moli\`ere's theory of multiple scattering}
\label{sec:distribution-ioe}
Adopting the small dipole size approximation, \eqn{eq:v_GW_text} (\eqn{eq:v_HTL_text}) can be expanded to linear order in $\mu^2\x^2$ ($m_D^2 \x^2$), which we refer to as the 
leading power (LP) contribution to $S(\x)$. The choice of this terminology is motivated by the desire of distinguishing between universal power corrections of the form $(Q_{s0}^2\x^2)^n$ (with $Q_{s0}^2\equiv \hat{q}_0 L$~\cite{IOE3}, see Appendix~\ref{appendix:dipole}) and non-perturbative power corrections of the form  $(\mu^2\x^2)^n$ ($(m_D^2\x^2)^n$), which we refer to as next-to-leading power terms. To leading power (LP), we obtain 
\beq 
v^{\rm GW}(\x)&=\dis\frac{\hat{q}_0}{4}\,\x^2 \log\left(\frac{4 \rme^{1-2\gamma_E} }{\x^2\mu^2} \right) +\cO(\x^4\mu^2)\,,
\eeq
and 
\beq 
v^{\rm HTL}(\x)&=\dis\frac{\hat{q}_0}{4}\, \x^2\log\left(\frac{4 \rme^{2-2\gamma_E} }{\x^2m_D^2} \right)+\cO(\x^4m_D^2)\,.
\eeq
We see that, at this level of accuracy, it is possible to relate the GW and HTL parameters by defining the following physical scale \footnote{We note that this map has, as far as we know, not been considered before. Rather, previous studies assumed $\mu=m_D$; see~\cite{Djordjevic:2007at}.},
\beq\label{eq:ir-map}
\mu_\ast^{2} =    \begin{dcases}
      \,   \frac{\mu^2}{4}\, \rme^{-1+2 \gamma_E} & \text{for the GW model}\\
      \,  \frac{m_{D}^2}{4}\, \rme^{-2+2 \gamma_E} & \text{for the HTL model}\\
    \end{dcases} \, .
\eeq
More concretely, this approximation leads to
\begin{align} \label{eq:S_LT}
 S^{\rm LP}(\x) = \exp\left[ -\frac{1}{4}Q^2_{s0} \,\x^2\,  \log \frac{1}{\x^2 \mu_\ast^{ 2}} \right]+\cO(\x^2\mu_\ast^{ 2})\,.
\end{align}
 Given the physical scale defined in \eqn{eq:ir-map}, this contribution is model independent. The possibility of expanding \eqn{eq:v_GW_text} and \eqn{eq:v_HTL_text} up to next-to-leading power order is explored in Section~\ref{sec:HT-GW}. To obtain an analytic form of $\cP(\k,L)$ from \eqn{eq:S_LT} additional assumptions concerning the parton's energy have to be adopted.

At very high energy, i.e. $\k^2 \gg Q^2_{s0}$, the dipole's transverse size is given by $|\x|\sim 1/|\k|\ll 1/Q_{s0}$. Then, $S^{\rm LP}(\x)$ can be expanded to linear order
\begin{align}
  S^{\rm LP}(\x)\Big\vert_{|\x|\ll 1/Q_{s0}} &= 1 -\frac{1}{4}Q^2_{s0} \,\x^2\,  \log \frac{1}{\x^2 \mu_\ast^{2}}+\mathcal{O}\left(\x^4Q_{s0}^4\right)\,.
\end{align}
The zeroth order term can be neglected as it does not contribute to the $\cP(\k,L)$ distribution. Thus, $S^{\rm LP}(\x)$ is proportional to the dipole cross-section, i.e. it is dominated by single hard (SH) scattering contributions. In this case, $\cP^{\rm SH}(\k,L)$ reads 
\beq\label{eq:P_SH}
 \cP^{\rm SH}(\k,L) &=& - \frac{1}{4}Q^2_{s0}  \int_\x \, \rme^{-i \x\cdot \k} \, \x^2 \log \frac{1}{\x^2 \mu_\ast^{ 2}} \nn
 &= &  \frac{1}{4}Q^2_{s0}\, \vec{\nabla}_\k^2  \frac{4\pi}{\k^2}   =4\pi  \frac{Q^2_{s0}}{\k^4}\,,
\eeq 
where used the fact that for $\k^2\gg \mu_\ast^2$
\beq
 \int_\x \,\rme^{-i \x\cdot \k} \log \frac{1}{\x^2 \mu_\ast^{2}}  = \frac{4 \pi }{\k^2}\,.
\eeq
The momentum broadening probability distribution given by \eqn{eq:P_SH} captures the expected Coulomb-like $1/\k^4$ behavior at high momentum transfers. 

The scale at which multiple scattering (MS) becomes important is encoded in the medium opacity parameter $\chi \sim Q_{0s}^2/\mu_\ast^{ 2} \sim L/\ell_{\rm mfp}$, where $\ell_{\rm mfp}$ is the in-medium mean free path. When $\chi \ll 1$, the medium is dilute and therefore single (rare) hard scattering events, as discussed above, dominate the contribution to $\cP(\k,L)$. Conversely, when $\chi \gg 1$, the medium is densely populated and multiple soft scatterings become the relevant mechanism for momentum transfer.\par
In the $\k^2 \ll Q_{s0}^2$ regime, the logarithm in \eqn{eq:S_LT} is slowly varying with $\x$ and can be regulated by a large momentum scale $Q^2 \sim Q_{s0}^2$, so that taking into account all orders in $Q_s^2\sim Q_{s0}^2 \log (Q^2/\mu^2)$ one obtains a Gaussian representation of momentum broadening
\beq\label{eq:P_MS}
 \cP^{\rm MS}(\k,L) &= &  \int_\x \,  \rme^{-\frac{1}{4} \x^2  Q_{s}^2 } \,\rme^{-i \x\cdot \k}=\frac{4 \pi }{Q_{s}^2}\, \rme^{-\frac{\k^2}{Q_{s}^2}}\,,
\eeq
where the super-script refers to multiple scattering (MS).
As mentioned in the introduction, \eqn{eq:P_MS} is the preferred option in widely used and successful jet quenching Monte Carlo event generators such as the Hybrid model~\cite{Hybrid} or the newly developed code by the Saclay group~\cite{Saclay}. Despite the widespread phenomenological application of \eqn{eq:P_MS}, it fails to accurately describe the hard $1/\k^4$ tail of $\cP(\k,L)$, thus missing the physics associated with single hard scattering events (see \eqn{eq:P_SH})~\footnote{An equivalent derivation of these two limits is given, in the context of the kinetic formulation of momentum broadening, in Appendix~\ref{appendix:kinetic_broad}}. The importance of such contributions has been recently studied numerically in~\cite{Kutak1,Kutak2}.

In this work, we propose to use an efficient expansion scheme developed by Moli\`ere in 1948 \cite{Moliere} in order to provide a simple, analytic formula that encodes the correct behavior of $\cP(\k,L)$ from small to large $\k$. Moli\`ere's approach is based on expanding $v(\x)$ around the multiple soft scattering approximation $v^{\rm{MS}}(\x)$. This is achieved by splitting the Coulomb logarithm into two pieces: a large, but constant, logarithm and a small, $\x^2$-dependent term which is treated perturbatively. Using this scheme, the leading power (LP) dipole cross-section can be written as
\begin{align}\label{eq:v_Moliere}
 v^{\rm LP}(\x)&=\frac{\hat{q}_0\x^2}{4}\log \frac{Q^2}{\mu_\ast^2}+\frac{\hat{q}_0\x^2}{4}\log \frac{1}{\x^2 Q^2}
 \nn
 &\equiv v^{\rm MS}(\x)+\delta v(\x)  \, ,
\end{align}
where $Q^2$ is known as matching scale, $v^{\rm MS}(\x)$ corresponds to the cross-section entering \eqn{eq:P_MS} and $\delta v(\x)$ can be considered a perturbation as long as $Q^2 \gg \mu_\ast^{ 2}$, such that $\log \frac{1}{\x^2 Q^2} \ll  \log \frac{Q^2}{\mu_\ast^{ 2}}$. This decomposition leads to the following definitions of the relevant scales in the problem
\beq\label{eq:Q-def}
Q_s^2 \equiv \langle \k^2 \rangle_{\rm typ} = \hat q_0 L  \,\log \frac{Q^2}{\mu_\ast^2}\,,
\eeq
where $ Q^2$ can be taken to be proportional to $Q_s^2$, i.e., 
\beq\label{eq:Q-Qs}
 Q^2 = a Q_s^2\,.
\eeq
Here $a$ is a free parameter to be determined for each set of medium parameters~\footnote{In the numerical results shown in this paper the prescription $a=1$ was taken}. Given a value of $a$, by inserting \eqn{eq:Q-Qs} in \eqn{eq:Q-def}, one obtains the following transcendental equation
\beq
\label{eq:Qs-def}
Q_s^2 = \hat q_0 L  \,\log \frac{aQ_s^2}{\mu_\ast^2}\, , 
\eeq
where the choice $a\!=\!1$ corresponds to Moil\`ere's prescription \cite{Moliere}. We also define the \textit{effective} transport coefficient $\hat{q}$ as 
\beq
 \hat q = \frac{\langle \k^2 \rangle_{\rm typ}}{L} =  \hat q_0   \,\log \frac{aQ^2_s}{\mu_\ast^2}\, .
\eeq
\par 
Using \eqn{eq:v_Moliere}, one can expand \eqn{eq:pt-broad-def} (at LP accuracy) around the MS solution in powers of $\delta v(\x)$ as

\begin{align}\label{eq:expli_cP_series}
\cP^{\rm LP}(\k,L)&=\sumint_{\x, \,n} \rme^{-i\x\cdot \k} \rme^{-\frac{1}{4}\x^2Q_s^2}\frac{(-1)^n Q_{s0}^{2n}}{4^nn!}\x^{2n} \log^{n}\frac{1}{\x^2Q^2} 
\nn
&\equiv \cP^{(0)}+\cP^{(1)}+\cP^{(2)} +\cdots \, ,
\end{align}
where we integrate over $\x$ and sum over $n$, from $n\!=\!0$ to infinity. Notice that we still have $\int_\k \cP^{\rm LP}(\k,L)\!=\!1$, since all terms in \eqn{eq:expli_cP_series}, apart from the (0) term, vanish after integrating over $\k$ and $\x$. Formally, the series representation introduced above  is asymptotically divergent. Nevertheless, a very good approximation of the exact solution can be obtained when the series is truncated before it diverges at $n < n_{\rm max}\sim Q_{s0}^2/\mu_\ast^2$. For a broader discussion on the origin of this truncation, we refer the reader to \cite{Iancu:2004bx}, where Moli\`ere's expansion was explored in the Color Glass Condensate framework.
\par
In order to recast \eqn{eq:expli_cP_series} in a more compact form we define the dimensionless expansion parameter
\beq \label{eq:lambda-def}
\lambda\equiv \frac{\hat{q}_0}{\hat{q}}=\frac{1}{\log(Q^2/\mu_\ast^{ 2})} \ll 1\, .
\eeq
This allows us to re-write \eqn{eq:expli_cP_series} as
\beq\label{eq:P_as_f}  
&& (4 \pi)^{-1} Q_s^2 \, \cP(\k,L)\equiv f\left(x,\lambda\right) = \sum_{n=0}^\infty \lambda^n f^{(n)} (x) \, ,
\eeq
where  $x\equiv \k^2/Q_s^2$.

The leading order (0) term in $\lambda$ reads 
\beq 
f^{(0)}=(4\pi)^{-1} \,Q_s^2\, I_1(x)= \rme^{-x}\,,
\eeq
while the next-to-leading order (1) term yields \cite{Moliere}
\beq
\label{eq:ioe-nlo}
\lambda f^{(1)} &&= - \frac{1}{16 \pi} Q^2_{s0} Q_s^2 \int_\x\,\rme^{-i \x\cdot \k}  \rme^{-\frac{1}{4} Q_s^2 \x^2} \, \x^2 \log \frac{1}{\x^2 Q^2} 
\nn
 && = \frac{Q_s^4}{16 \pi} \lambda  \vec{\nabla}^2_\k\, \int_\x\,\rme^{-i \x\cdot \k}  \rme^{-\frac{1}{4} Q_s^2 \x^2} \,\log \frac{1}{\x^2 Q^2} 
 \nn
 &&= \frac{\lambda Q_s^2}{4\pi}\,  \frac{\del }{\del x } x\frac{\del }{\del x } \, I_2(x,a)\, \nn
 &&= \lambda\, \Delta_x \,e^{-x}\, \left({\rm Ei}\left(x\right)-\log (4x\,a)\right)\,,
\eeq 
where $I_1(x)$ and $I_2(x,a)$ are given in Appendix~\ref{appendix:integrals} and Ei is the exponential integral function~\footnote{Here we used $\vec{\nabla}^2_\k=4/Q_s^2 \Delta_x$ with $x\equiv \k^2/Q_s^2$}. Here we introduced the reduced Laplacian operator $\Delta_x\equiv\partial_x(x\,\partial_x)$. Combining the results for $f^{(0)} $ and $f^{(1)} $
, we obtain one of the main results of this paper (originally derived by Moli\`ere \cite{Moliere}):  
\begin{widetext}
\begin{align}
\label{eq:golden}
    \cP^{(0)+(1)}(\k,L)=\frac{4\pi}{Q_s^2} \rme^{-x}\Big
 \{1 -\lambda \left(\rme^{x}-2+\left(1-x\right)\,\big({\rm Ei}\left(x\right)-\log (4x\,a)\right) \big)\Big\} \, , \quad x\equiv\frac{\k^2}{Q_s^2} \, .
\end{align}
\pagebreak 
\end{widetext}
Let us now verify that the above result reproduces the expected asymptotic behavior at small and large transverse momentum.\par  
The (0) contribution matches exactly the MS solution from \eqn{eq:P_MS}, and thus its limiting behavior is easy to analyse. At large momentum transfers, $\k^2\gg Q_s^2$, $\cP^{(0)}(\k,L)$ decays exponentially with $\k$, while if $\k^2\ll Q_s^2$ it becomes independent of $\k$. More importantly, the Gaussian profile implies that the typical momentum transverse acquired due to momentum broadening $\langle\k^2\rangle_{\rm typ}\!\sim\! Q_s^2$. Therefore, $\cP^{(0)}(\k,L)$ correctly captures the physics associated to multiple soft scattering at scales $\k^2\lesssim Q_s^2$.\par 
For the (1) term, we use two limiting forms of the Ei function. That is, when $x\to\infty$, ${\rm Ei}(x)\approx \rme^x (1/x+1/x^2+2/x^3)$ so that the large $\k$ behavior of the (1) term reads
\begin{align}
\cP^{(1)}(\k,L)\Big\vert_{\k^2\gg Q_{s}^2 }=4\pi\frac{Q_{s0}^2}{\k^4}+\mathcal{O}\left(\frac{Q_{s0}^4}{\k^6}\right) \, .
\end{align}
This result matches \eqn{eq:P_SH} and therefore the (1) term successfully encodes the hard $1/\k^4$ tail of the full $\cP(\k,L)$ distribution. As a consequence, it is physically preferable for phenomenological applications to use \eqn{eq:golden} instead of \eqn{eq:P_MS}. On the other end, $x\to0$, ${\rm Ei}(x)\approx \gamma_E+\log x$ and then 
\begin{align}
\cP^{(1)}(\k,L)\Big\vert_{\k^2\ll Q_{s}^2 }= \frac{4\pi\lambda}{Q_s^2}\log 4\, a\, \rme^{1-\gamma_E} \, ,
\end{align}
which, up to a small constant logarithm, corresponds to the MS result (\eqn{eq:P_MS}) in this kinematic limit, suppressed by a power of $\lambda$. This is analogous to the small energy limit behavior obtained in \cite{IOE1,IOE2,IOE3} for the gluon emission spectrum.\par

\begin{figure}
\centering
\includegraphics[width=\columnwidth]{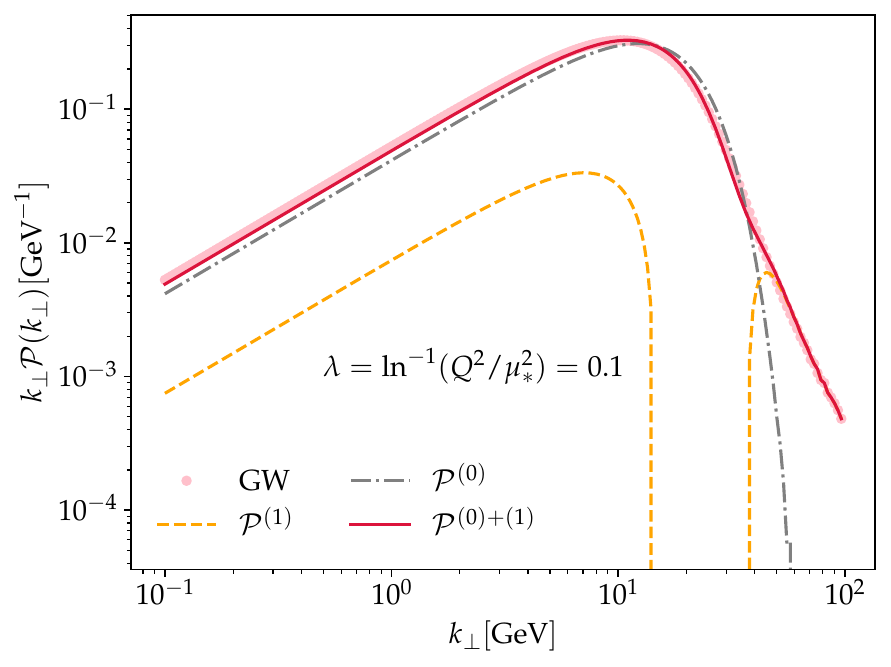}
\caption{Momentum broadening probability distribution at different orders in Moli\`ere's-expansion (see \eqn{eq:expli_cP_series}) compared to the exact result for GW (see \eqn{eq:v_GW_text}) with $\lambda\!=\!0.1$ corresponding to ($Q^2_{s0}\!=\!30$~GeV$^2$, $m^2_D\!=\!0.13$~GeV$^2$). In this and following figures $k_T\equiv |\k|$.}
\label{fig:theory-plot}
\end{figure}

In Fig.~\ref{fig:theory-plot}, we evaluate \eqn{eq:P_as_f} at (0), (1) order and their sum (0)+(1). These curves are compared to the exact numerical solution of \eqn{eq:pt-broad-def} when plugging the GW dipole cross-section given by \eqn{eq:v_GW_text}. A small value of the expansion parameter $\lambda$ is  chosen on purpose such that this figure represents a proof-of-concept of the proposed scheme in its regime of validity. In the multiple scattering regime, i.e. at small-$k_\perp$, the (0) contribution dominates over the (1) term as expected from our asymptotic analysis. Nevertheless, the (0)+(1) curve shows a small discrepancy, to be quantified in what follows, with respect to the full GW result. The situation is improved at large-$k_\perp$, where the (1) contribution correctly captures the power-law tail completely absent in the (0) scenario. This figure demonstrates how a purely analytic, two terms expansion given by \eqn{eq:golden} exhibits an excellent agreement with the numerically obtained $\cP(\k,L)$ using GW/HTL models for $v(\x)$.   

\begin{figure}
\centering
\includegraphics[width=\columnwidth]{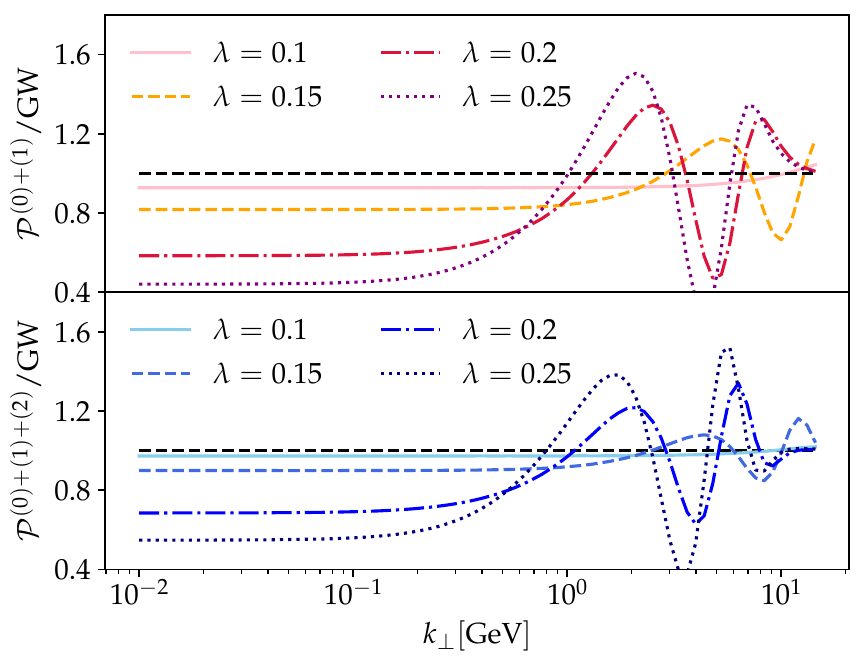}
\caption{Top: ratio between the (0)+(1) result and the exact GW varying the expansion parameter $\lambda$ (see \eqn{eq:lambda-def}). Bottom: same as top panel but for the (0)+(1)+(2) result. $\lambda\!=\!0.15,0.2,0.25$ corresponds to ($Q^2_{s0}\!=\!4$~GeV$^2$, $m^2_D\!=\!0.3$~$\rm GeV^2$), ($Q^2_{s0}\!=\!4$~$\rm GeV^2$, $m^2_D\!=\!0.5$~$\rm GeV^2$), and ($Q^2_{s0}\!=\!1.5$~$\rm GeV^2$, $m^2_D\!=\!1$~$\rm GeV^2$) respectively.}
\label{fig:ioe-lambda}
\end{figure}

The natural question arises as to what is the value of the $\lambda$-parameter at which the expansion fails to reproduce the GW result. This problem, together with the role of higher orders, is tackled in Fig.~\ref{fig:ioe-lambda}. In the top panel we observe how the performance of the (0)+(1) truncation is degraded when increasing $\lambda$ both at low and large $\k$. This result is expected as the larger $\lambda$ gets, the less precise is to consider $\delta v(\x)$ as perturbative contribution in \eqn{eq:v_Moliere}. The relevant values of $\lambda$ for current and future colliders will be discussed in Section~\ref{sec:numerics}. As shown in the bottom panel of Fig.~\ref{fig:ioe-lambda} this discrepancy can be alleviated by adding extra terms in the expansion. In particular, when adding the (2) (see \eqn{eq:expli_cP_series}) contribution for $\lambda\!=\!0.1$ we find a ratio to the exact GW result close to one in the whole interval in $\k$. Unfortunately, we were not able to find yet a general analytic expression for the $n$-th term in the series, thus higher orders have to be computed numerically. 

\section{Sensitivity to IR modeling up to next-to-leading power order REMOVE[via a twist expansion]: GW and HTL comparison }
\label{sec:HT-GW}
In the previous Section, as a first step towards an analytic expression for $\cP(\k,L)$ that encompasses the main physical mechanisms, we have expanded the dipole cross-section to leading power order (see \eqn{eq:S_LT}).
In order to assess the sensitivity of transverse momentum broadening to the non-perturbative infrared structure of a given model for $\gamma(\q)$, going beyond the leading power (LP) term is mandatory. In what follows, we fix the hard scale of the problem $Q_{s0}$ such that we are only sensitive to the dependence of this expansion with respect to the infrared regulator $\mu_\ast$. Consequently, these additional terms are expected to modify the low momentum regime of $\cP(\k,L)$. Hence, this will be the explored region in the figures of this section. At this point, we would like to emphasize that the expansion in universal and non-universal terms is intrinsically different from the Moli\`ere one introduced in the previous section. In short, the former explores the infrared sector through non-universal contributions, while the latter is a perturbative expansion with model independent terms. 

A systematic study of these non-universal contributions is so far missing in the literature. In turn, GW and HTL models are typically treated as two independent descriptions of the QCD medium, thus ignoring the fact that they can be mapped onto one another at LP accuracy. The common practice in jet quenching phenomenology of treating these models as if they were unrelated can be problematic given that: i) in the absence of a map between the different IR regulators to the physical Debye mass, any comparison between results assuming different models is meaningless, ii) a quantitative and controlled understanding of the role of non-perturbative physics at the infrared scale cannot be reached. \par
\begin{figure}
\centering
\includegraphics[width=\columnwidth]{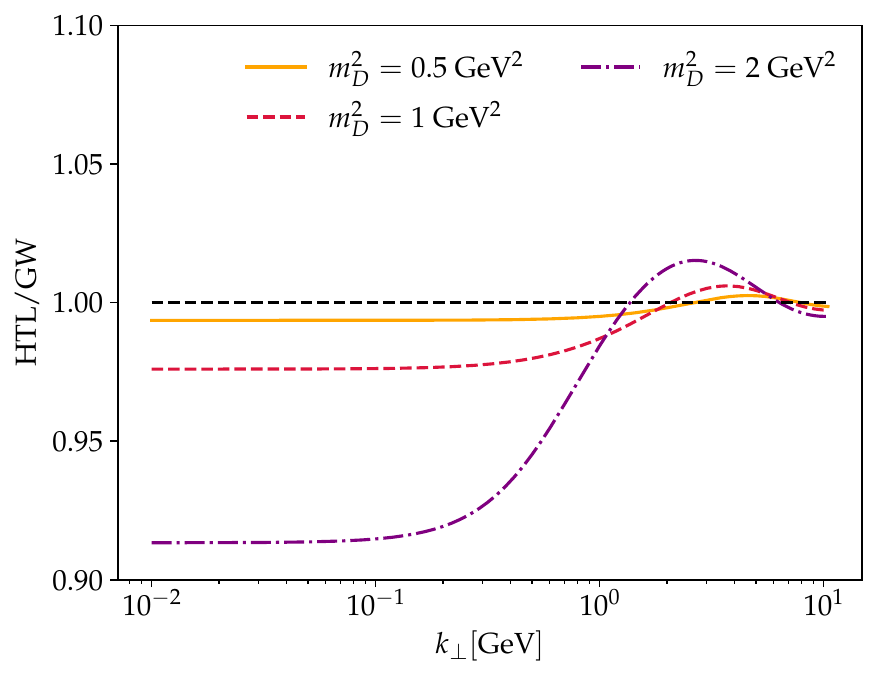}
\caption{Ratio between $\cP^{\rm HTL}(\k,L)$ and $\cP^{\rm GW}(\k,L)$ as a function of the Debye mass $m^2_D$ for $Q^2_{s0}\!=\!4.8$~GeV$^2$ (same value used in the other plots of this Section).}
\label{fig:gw-htl}
\end{figure}

 
The importance of mapping the IR scales involved in GW and HTL becomes apparent in Fig.~\ref{fig:gw-htl} where the value of $m_D$ is fixed and the corresponding value for $\mu$, to be plugged in \eqn{eq:v_GW_text}, is obtained using \eqn{eq:ir-map}. The momentum broadening probability distribution computed with these two medium models only shows significant differences at small-$k_\perp$ for values of $m^2_D$ larger than $\sim1\,{\rm GeV}^2$.
Hence, once the appropriate matching between $m_D$ and $\mu$ is considered, the use of these two in-medium elastic scattering cross-sections leads to discrepancies $\leq 10\%$ in the infrared sector~\footnote{We would like to remind the reader that, for any practical application, the contribution coming from the the region $|\k|\ll Q_{s0}$ is small, since it corresponds to the description of large dipole sizes, which play a secondary role when describing in-medium scattering}. This result is in agreement with previous calculations where the mapping was used~\cite{IOE3,CarlotaFabioLiliana}.

To characterize the sensitivity of transverse momentum broadening to infrared physics, we proceed to expand $v(\x)$ as given by \eqn{eq:v_GW_text} (\eqn{eq:v_HTL_text}) up to next-to-leading power (NLP) order, such that $v(\x)$ takes the form
\begin{align}\label{eq_v_NLT}
v^{\rm LP+NLP}(\x)&=\frac{\hat{q}_0\x^2}{4}\log\left(\frac{1}{\mu^2_\ast\x^2}\right)+\frac{\hat{q}_0\x^4\mu_\ast^2}{c_1}\log\left(\frac{c_2}{\mu_\ast^4\x^4}\right) 
\nn
&\equiv v^{\rm LP}(\x)+v^{\rm NLP}(\x)
\, ,
\end{align}
where $c_1$ and $c_2$ are model dependent constants given in Appendix~\ref{appendix:models} for the GW and HTL models. We would like to point out that \eqn{eq_v_NLT} is an explicit manifestation of the fact that NLP corrections are non-universal, since there is no self consistent way of mapping different models to some functional form representing what would be measured in experiment. In fact, even using the LP map introduced in \eqn{eq:ir-map}, we clearly observe that universality is not regained (i.e. $c_1$ and $c_2$ are truly model dependent). Nonetheless, using such a map allows us to directly gauge the order of magnitude of the model dependent contributions to $\cP(\k,L)$. \par 

\begin{figure}
\centering
\includegraphics[width=\columnwidth]{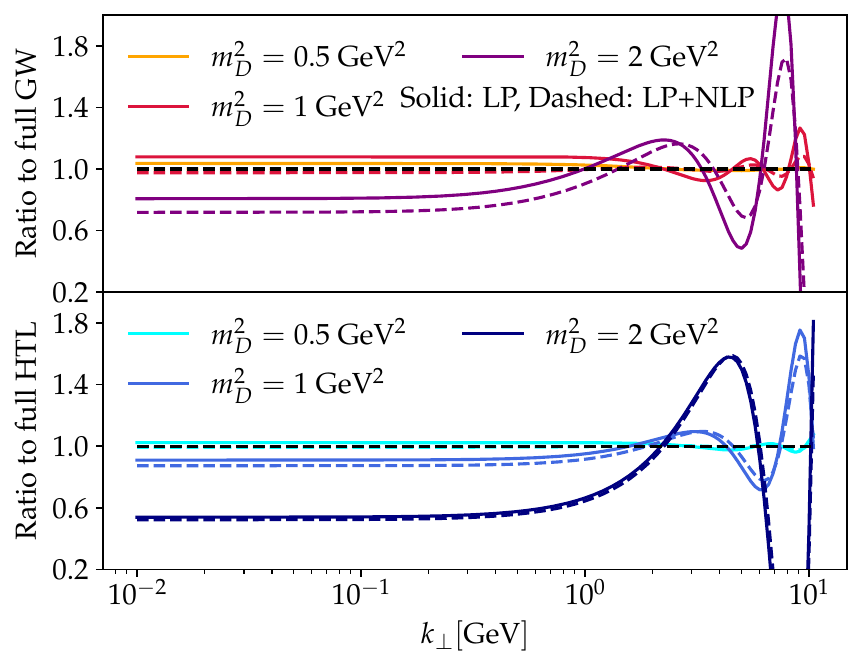}
\caption{Ratio of the leading power (LP) (solid) and next-to-leading power (NLP) expansions to the full GW (top) and HTL (bottom) potentials for different $m^2_D$. Notice that the orange, dashed line in the top panel fully overlaps with the reference black line.}
\label{fig:gw-htl-twists}
\end{figure}

The implications of considering the LP or NLP approximations to GW/HTL potentials instead of the full result are shown in Fig.~\ref{fig:gw-htl-twists}. Regarding the leading power term, it fails to reproduce the full result accurately for relatively small values of $m_D$, similar to those explored in current colliders as will be discussed in the next Section. Two comments are in order concerning the next-to-leading power term. First, focusing on the HTL case, its magnitude is remarkably small. This is an important aspect as it indicates a close to minimal sensitivity to non-perturbative assumptions. Second, when turning to the GW case this difference with respect to the LP is significantly more pronounced. The underlying reason is that the map between $m_D$ and $\mu$ through the physical mass $\mu_\ast$ is valid only at LP order. Therefore, when extending its applicability to NLP, more severe deviations between GW and HTL are deemed to occur.

Further, it is possible to combine the next-to-leading power expansion of the dipole cross-section (see \eqn{eq_v_NLT}) with Moli\`ere's scheme, by shifting the expansion point $v^{\rm MS}(\x)\rightarrow v^{\rm MS}(\x)+v^{\rm NLP}(\x)$ and continue treating $\delta v(\x)$ (see \eqn{eq:v_Moliere}) as a perturbation around it. In this way, the $\lambda$-expansion enables the description of $\cP$ for all $\k$, while it is possible to explore the dependence on IR modeling via the expansion beyond the LP contribution. 

The first non-trivial term in \eqn{eq:expli_cP_series} is thus promoted (to linear order in $\mu^2_\ast$~\footnote{The deviation between \eqn{eq:analy_exp} and the full NLP broadening probability $\cP^{\rm (1)+NLP}(\k,L)$ obtained combining \eqn{eq_v_NLT} and Moli\`ere's scheme is numerically and parametrically negligible. This is due to the fact that corrections to the spectrum coming from Moli\`ere's theory are dominant over the corrections due to non-perturbative modeling. \eqn{eq:analy_exp} is thus preferred since it can be written in a closed analytical form.}) to the following simple expression,
\begin{align}\label{eq:analy_exp}
\cP^{(1)+\delta NLP}(\k,L)&= \lambda \Delta_x \, I_2(x,a)
\nn
&-\frac{32\lambda \mu^2_\ast}{c_1Q_s^2}\Delta_x^2\, I_2\left(x,\frac{\mu_\ast^2}{\sqrt{c_2} Q_s^2}\right)\, .
\end{align}
Equation~\eqref{eq:analy_exp} is no longer a function of solely $Q^2/\mu_\ast^2$ (i.e. $\lambda$), but also depends on the IR regulator alone and on the constants $c_1$ and $c_2$; this is an explicit example of the non-universality attribute we have associated to these terms.

\begin{figure}
\centering
\includegraphics[width=\columnwidth]{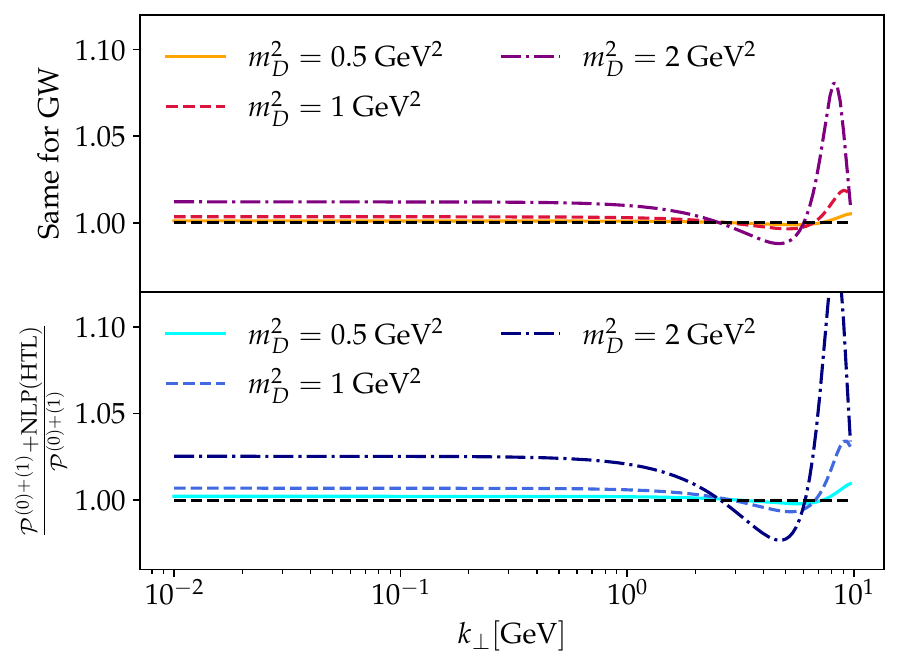}
\caption{Impact of the NLP term in Moli\`ere's scheme at (1) order (using \eqn{eq_v_NLT}) for GW (top) and HTL (bottom) potentials for different $m^2_D$.}
\label{fig:ioe-twist}
\end{figure}

The role of the NLP contribution in Moli\`ere's expansion at first (1) order is displayed in Fig.~\ref{fig:ioe-twist} for both GW (top) and HTL (bottom) potentials. In all cases, the inclusion of the NLP term has an effect smaller than $5\%$ being this contribution larger when increasing $m_D$, as expected. This fact confirms the mild infrared dependence in our description of $\cP(\k,L)$. Further, the NLP contribution enhances the value of $\cP(\k,L)$ at small-$k_\perp$ thus reducing the discrepancy with the full GW/HTL result observed in Figs.~\ref{fig:theory-plot} and \ref{fig:ioe-lambda}.

\section{Relevance for phenomenology at current and future colliders}
\label{sec:numerics}

Up to now, we have performed a theory guided selection of the parameters involved in the description of the medium in order to highlight the relevant region of interest. That is, in Section~\ref{sec:distribution-ioe} we have deliberately chosen small values of $\lambda$ as required by Moli\`ere's scheme or large values of $m_D$ in Section~\ref{sec:HT-GW} to better illustrate the contribution of next-to-leading power terms.

In what follows, we turn our attention to more realistic scenarios such as the ones being currently explored by LHC and RHIC together with the upcoming EIC. The relevant parameters for these three colliders are provided in Table~\ref{table:nonlin}. For a given value of the medium length, $L$, and its temperature, all parameters in this table are uniquely determined. More concretely, $m_D$ is obtained through the leading order HTL formula mentioned in Section~\ref{sec:kinetic} and $Q_s^2$ is given by \eqn{eq:Qs-def} where $\hat q_0\!=\!18 \pi \alpha_s^2T^3$, where we take $\alpha_s=1/\pi$. In all three cases, we take the length of the medium to be $L\!=\!5$~fm roughly corresponding to the radius of both Pb and Au nuclei. For RHIC and LHC, we use the temperature estimates $T_{\rm RHIC}=220 \, {\rm MeV}$ and $T_{\rm LHC}=300 \, {\rm MeV}$, corresponding to an indicative  average temperature for each setup in accordance with the values found in e.g.~\cite{Feal:2019xfl,new_exp_v_1,new_exp_v_2}. We also assume the medium can be described by the HTL model. 
Turning to the EIC case, the characteristic scale of the medium explored by the partonic probe is no longer the temperature, but the nucleon size that can be approximated as $1/\Lambda_{\rm QCD}$. This provides a natural value for the infra-red regulator in the GW model, i.e. $\mu=\Lambda_{\rm QCD}=200 \, {\rm MeV}$. Regarding $\hat{q}_0 L$, we make use of the Color Glass Condensate effective field theory to estimate its value following~\cite{Kolbe:2020tlq}, leading to $\hat{q}_0 L = 0.35 \, {\rm GeV}^2$. Note that this is in the ballpark of the value used in ~\cite{Liu:2018trl}. We consider in what follows quark jets with $C_R=C_F=\frac{4}{3}$.

\begin{table}[ht]
\caption{Relevant parameters for the three different setups.}
\centering
\begin{tabular}{|c| c |c| c| c|c|}
\hline
Collider & $T$[MeV] & $m_D^2$ [GeV$^2$] & $Q_s^2$[GeV$^2$] & $\hat q$ [GeV$^2$/fm] & $\lambda$ \\ [0.5ex] 
\hline
LHC &300&0.54&23.16&4.63&0.17\\
RHIC &220&0.29&8.56&1.71&0.18\\
\hline
Collider & $\mu$[MeV] & $\mu^2_\ast$[GeV$^2$] & $Q_s^2$[GeV$^2$] & $\hat q$ [GeV$^2$/fm] & $\lambda$ \\ [0.5ex] 
\hline
EIC &200&0.01& 1.75&0.35&0.2\\
\hline
\end{tabular}
\label{table:nonlin}
\end{table}

\begin{figure}
\centering
\includegraphics[width=\columnwidth]{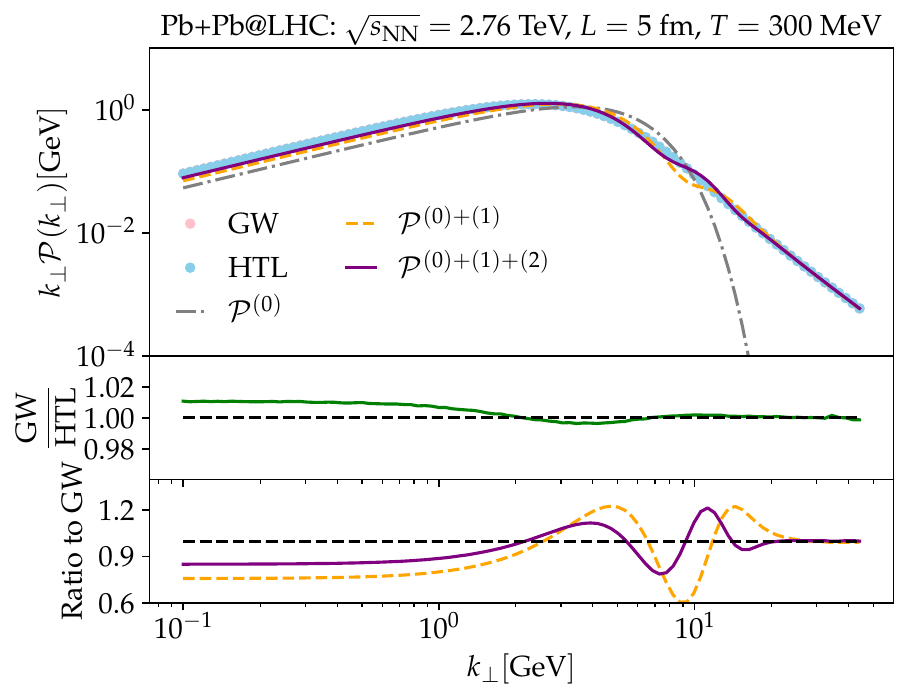}
\caption{Top: Momentum broadening probability distribution at different orders in Moli\`ere's expansion, see \eqn{eq:expli_cP_series}, compared to the exact result for GW and HTL potentials at LHC conditions. Middle: Ratio between GW and HTL results. Bottom: Ratio between (0)+(1) (orange) and (0)+(1)+(2) (purple) truncations with respect to the GW result.}
\label{fig:lhc-results}
\end{figure}

The results for Pb+Pb collisions at LHC energies are shown in Fig.~\ref{fig:lhc-results}. Regarding the large-$k_\perp$ region, the inability of the leading order term to capture the behavior of both GW and HTL models becomes apparent in the top panel. In contrast, the (0)+(1) and (0)+(1)+(2) contributions deliver a distribution identical to the GW result up to 5\% degree of accuracy for $k_\perp>\!20$~GeV. This statement also applies to HTL because the difference among these models is negligible in this region, as shown in the middle panel. In the infrared sector, $k_\perp<1$~GeV, we observe deviations of order $20 \%$ which can be improved with higher order terms, although this sector's contribution to physical observables is typically negligible. Around the peak of the distribution, the disagreement between (0)+(1)+(2) order with GW is of the order of 20\%. This result could be improved by adding higher order terms in Moli\`ere's expansion (see \eqn{eq:expli_cP_series}).

\begin{figure}
\centering
\includegraphics[width=\columnwidth]{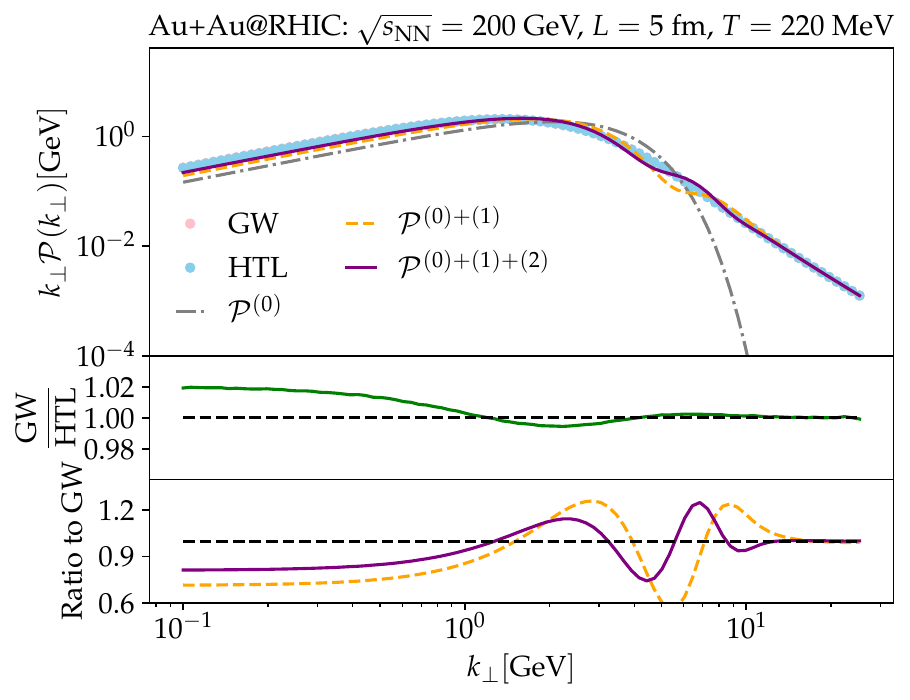}
\caption{Same as Fig.~\ref{fig:lhc-results} but for RHIC.}
\label{fig:rhic-results}
\end{figure}

Turning to RHIC energies, similar features to those at LHC are observed at large transverse momentum as shown in Fig.~\ref{fig:rhic-results}. An important remark concerning the small-$k_\perp$ sector is a slight increment of the GW vs. HTL disparity with respect to the LHC case, despite probing a smaller $m_D$, see Tab.~\ref{table:nonlin}. The underlying reason is that the value of $m_D$ has to be sufficiently separated from the hard scale in the problem. That is, the larger the value of the ratio $Q_{s0}/m_D$ is, the smaller the sensitivity to the infrared, model-dependent contributions. Indeed, this ratio is 20\% bigger at LHC than RHIC, thus explaining the mild differences between GW and HTL for these two experiments.

\begin{figure}
\centering
\includegraphics[width=\columnwidth]{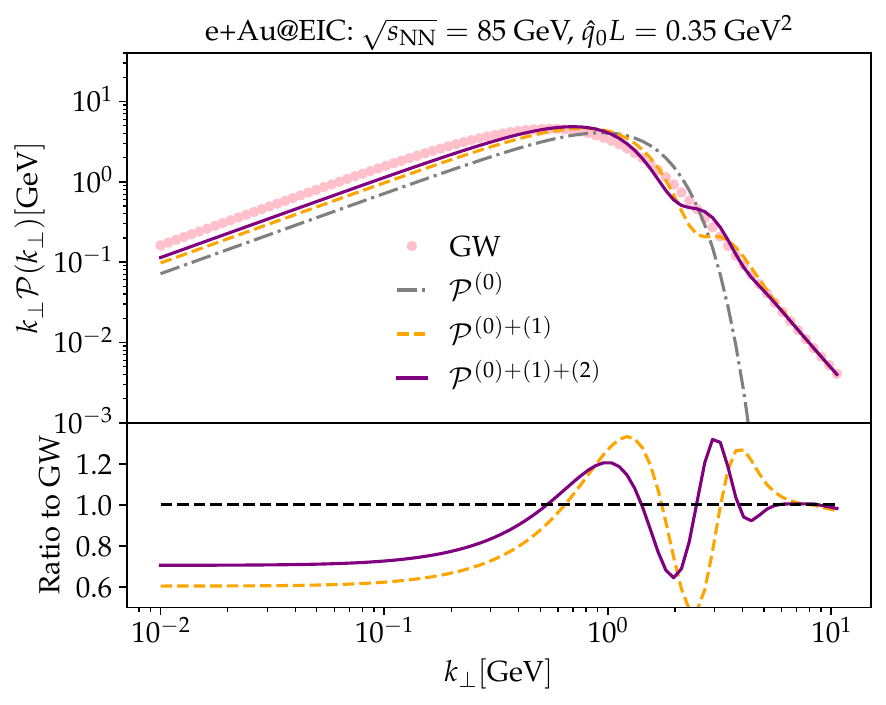}
\caption{Same as Fig.~\ref{fig:lhc-results} but for the future EIC. Note that we do not consider the HTL potential as a thermal medium at the EIC is not expected.}
\label{fig:eic-results}
\end{figure}

Finally, the future Electron-Ion Collider at Brookhaven National Lab~\cite{EIC} will open a new avenue to study modifications in jet observables when compared to hadronic colliders. The reason is that highly-energetic partons will not encounter a thermalized QGP at the EIC, but rather a dense gluonic system. Multiple interactions with this over-occupied gluon state naturally leads to transverse momentum broadening. Its probability distribution is displayed in Fig.~\ref{fig:eic-results}, where we have assumed e+Au to be the collision system, and the prospected top energy for this machine and the GW model is employed. We would like to point out that, in this case, we obtain that the relevant hard scale is $Q_s^2\!=\!1.75$~GeV$^2$.
At the same time, the average transverse momentum received from the \textit{cold} medium is shifted towards smaller values when compared to RHIC and LHC. This experiment will probe the largest value of $\lambda$ and, therefore, the application of Moli\`ere's scheme is less successful than for the LHC and RHIC setups. Nevertheless, the evolution of $\lambda$ as one changes the experimental set-up is slow and therefore this method is suitable for semi-quantitative exploratory studies.

The role played by the non-perturbative contributions at RHIC, LHC and EIC energies its suppressed when compared to the deviations due to the (1) expansion, as was shown in the previous Section. Therefore, we refrain from providing the numerical results obtained when considering these contributions.

\section{Conclusions and Outlook}
\label{sec:summary}
The work presented in this manuscript follows the current (and future) global effort towards a more precise quantitative description of jet quenching effects~\cite{IOE1,IOE2,IOE3,CarlotaFabioLiliana,Feal:2019xfl}. In particular, we have: i) re-derived a description of the single particle momentum broadening distribution, first introduced by Moli\`ere, that is able to reproduce the Gaussian behavior at small-$k_\perp$ together with the power-law tail through a simple analytic expression, i.e. \eqn{eq:golden}; ii) explored the impact of non-perturbative modeling of the dipole cross-section on the broadening distribution; iii) studied the applicability of Moli\`ere's scattering theory for the description of single particle momentum broadening at the current experimental conditions explored by LHC, RHIC and the foreseen EIC.\par 
Although the major analytic result of this paper (i.e. \eqn{eq:golden}) has been known for over 70 years, its application to phenomenological studies in jet quenching has, to a large extent, been ignored. This is particularly surprising given the widespread usage of the multiple soft scattering parametrisation (i.e. \eqn{eq:P_MS}) that is known to fail to describe the hard sector, as we have shown both analytically and numerically for meaningful parameter selections. It should be noted that Moli\`ere's proposal, if restricted only to the first two leading terms whose expressions are known analytically, requires a pronounced hierarchy between the hard and soft scales of the problem in order to converge to the exact result when considering GW/HTL potentials. In particular, we found that such a separation is reasonably satisfied at current LHC and RHIC conditions, but is far from being achieved at the expected EIC setup. \par 
Another important finding of this work is that non-perturbative contributions, intrinsically associated to the description of the infrared structure of in-medium scattering, are found to be quantitatively small. However, we (again) emphasize that to achieve a meaningful comparison between medium models, matching to a physical set of parameters at leading power (LP) accuracy is required. For the three physically interesting cases explored in Section~\ref{sec:numerics}, non-perturbative contributions are highly suppressed compared to the ones obtained from Moli\`ere's scheme. \par 
The present results are of relevance for jet quenching phenomenology and have, for example, been recently applied to computing the nuclear modification factor $R_{\rm{AA}}$~\cite{Mehtar-Tani:2021fud}. Interesting future studies could focus on the impact of the single hard scattering tail on jet substructure observables such as the groomed opening angle~\cite{Ringer:2019rfk} or the transverse momentum of the hardest splitting in the jet tree~\cite{Dyg}, and be further related to the presence of Moli\`ere scattering centers in the Quark Gluon Plasma~\cite{DEramo:2018eoy,DEramo:2012uzl}.

As mentioned in Section \ref{sec:distribution-ioe}, \eqn{eq:golden} can be readily employed in any Monte Carlo event generator that includes jet quenching effects. In addition, the formalism described in this paper can also be applied to small-$x$ physics \cite{final1}. A possibility would be to use Moli\`ere's scheme to construct the unintegrated gluon distribution needed for particle production in p+A~\cite{Kharzeev,Iancu:2004bx}.

\section*{ACKNOWLEDGMENTS}
We are grateful to Xabier Feal for pointing out the generalized form used in \eqn{eq:pt-broad-def} and for many other useful discussions. The project that gave rise to these results received the support of a fellowship from ``la Caixa" Foundation (ID 100010434). The fellowship code is LCF/BQ/ DI18/11660057. This project has received funding from the European Union's Horizon 2020 research and innovation program under the Marie Sklodowska-Curie grant agreement No. 713673. J.B. is supported by Ministerio de Ciencia e Innovacion of Spain under project FPA2017-83814-P; Unidad de Excelencia Maria de Maetzu under project MDM-2016-0692; European research Council project ERC-2018-ADG-835105 YoctoLHC; and Xunta de Galicia (Conselleria de Educacion) and FEDER. The work of Y. M.-T. and A. S.-O. was supported by the U.S. Department of Energy, Office of Science, Office of Nuclear Physics, under contract No. DE- SC0012704, and by Laboratory Directed Research and Development (LDRD) funds from Brookhaven Science Associates. K. T. is supported by a Starting Grant from Trond Mohn Foundation (BFS2018REK01) and the University of Bergen. Y. M.-T. acknowledges support from the RHIC Physics Fellow Program of the RIKEN BNL Research Center.

\appendix
\section{Typical integrals appearing at (1) order in Moli\`ere's theory}
\label{appendix:integrals}
In this Appendix we compute the following Fourier transforms
\beq\label{eq:I1}
I_1=  \int_\x \rme^{-i \x\cdot\k}\, \rme^{-\frac{1}{4}\,Q_s^2\, \x^2 \, } \, ,
\eeq
and 
\beq\label{eq:I2}
I_2=  \int_\x \,\rme^{-i \x\cdot\k}\, \,\rme^{-\frac{1}{4}\,Q_s^2\, \x^2 \, } \log \frac{1}{Q^2\x^2}\, .
\eeq
\eqn{eq:I1} is a straightforward to compute ($x \equiv\k^2/Q_s^2$)
\beq\label{eq:I1-res}
I_1(x) =  \frac{4\pi}{Q_s^2} \, \rme^{-x}\,.
\eeq
\eqn{eq:I2} makes use of the following integral representation of the logarithm
\beq
\log\frac{1}{\x^2Q^2} = -\lim_{\epsilon \to 0}\int_\epsilon^\infty \frac{\rmd t}{t} \left(\rme^{- t }-\rme^{-\x^2 Q^2 t }\right)\,.
\eeq
$I_2$ can then be rewritten in terms of two Gaussian integrals (using $a=Q^2/Q_s^2$)
\beq\label{eq:I2-form1}
 I_2&&= - I_1\, \int_{\epsilon}^\infty \frac{\rmd t}{t}\rme^{-t}
 + \int_{\epsilon}^\infty \frac{\rmd t}{t}  \int_\x\,\rme^{-i \x\cdot\k} \rme^{-\frac{1}{4}\left(\,1\,+4at\right) Q_s^2\x^2 \, } \nn
 &&= - I_1\, \int_{\epsilon}^\infty \frac{\rmd t}{t}\rme^{-t}  +\frac{4\pi}{Q^2} \int_{\epsilon}^\infty \frac{\rmd t}{t (1\,+4at)}  \,\rme^{-\frac{\k^2}{(1+4at)Q_s^2}}\,.\nn
 \eeq
Performing the change of variables $ u+x=x/(1+4at)$, the last integral in \eqn{eq:I2-form1} yields
\beq\label{eq:I2-second-term}
  -\rme^{-x}\int^{-4 ax \epsilon}_{-x} \frac{\rmd u}{u} \rme^{-u}=\,\rme^{-x} \left[ {\rm Ei}(x) - {\rm Ei}(4 ax \epsilon) \right]\nn
\eeq
Taking $\epsilon\to 0$, the first term in \eqn{eq:I2-form1} and the last term in \eqn{eq:I2-second-term} (after factoring out $I_1$) combine to give
\beq
&&  -\int_{\epsilon}^\infty \frac{\rmd t}{t}\rme^{-t} - {\rm Ei}(4 ax \epsilon) =  {\rm Ei}(\epsilon) - {\rm Ei}(4a x \epsilon) \nn
  &&= - \log 4ax\,+\cO(\epsilon)\, ,
\eeq
where we used that ${\rm Ei}(\epsilon)\simeq \gamma_E+\log \epsilon$, with $\gamma_E=0.577(2)$ the Euler-Mascheroni constant. Finally, we obtain 
\beq
I_2(x,a) = I_1(x) \Big[ {\rm Ei}(x)-\log 4a x \Big] \,.
\eeq

\section{Dipole cross-section in the GW and HTL models}
\label{appendix:dipole}
In this Appendix we compute the following integral 
\begin{equation}
\int_0^\infty du \ \frac{u}{(u^2+b^2)\left(u^2+a^2\right)}\left(1-J_0(ux)\right) \,,
\end{equation}
which is related to the GW and HTL models by letting $b=a=\mu$ and $b=0$, $a=m_{\rm D}$, respectively (see \eqn{eq:HTL}, \eqn{eq:GW}  and \eqn{eq:v}).  
First we decompose the integrand as follows
\beq
&& \int_0^\infty du \ \frac{u}{(u^2+b^2)\left(u^2+a^2\right)}\left(1-J_0(ux)\right)=\nn
&& \frac{1}{(a^2-b^2)} \int_0^\infty du  \ \left[\frac{u}{(u^2+b^2)}-\frac{u}{ \left(u^2+a^2\right)}\right]\left(1-J_0(ux)\right)\,.\nn
 \eeq
Recognizing the following integral representation of Bessel functions 
\beq
  \int_0^\infty du  \ \left[\frac{u}{(u^2+a^2)}\right]J_0(xu) = K_0(a x) 
 \eeq
 and 
 \beq
 \int_0^\infty du \ \frac{u}{(u^2+b^2)\left(u^2+a^2\right)} = \frac{\log a^2-\log b^2}{2(a^2-b^2)}\,,
 \eeq
 we obtain 
\beq
&& \int_0^\infty du \ \frac{u}{(u^2+b^2)\left(u^2+a^2\right)}\left(1-J_0(ux)\right)\nn
&&=\frac{1}{(a^2-b^2)} \left[ K_0(ax) -K_0(bx)+\log a -\log b \right]\,. \nn
\eeq
There are two special cases that will correspond to the two models under consideration in the main text. First, $a=b$  
 \beq
&& \int_0^\infty du \ \frac{u}{(u^2+a^2)^2}\left(1-J_0(ux)\right)\nn
&&=\frac{1}{2 a^2 } \left[ 1 -  a x  K_1(ax) \right]\,. \nn
\eeq
Then for $b=0$, using the form $K_0(bx) \approx - \log(b x/2) -\gamma_E$ 
\beq
&& \int_0^\infty du \ \frac{1}{u\left(u^2+a^2\right)}\left(1-J_0(ux)\right)\nn
&&=\frac{1}{a^2} \left[ K_0(ax) + \log(a x/2) +\gamma_E \right]\,. \nn
\eeq

\section{GW and HTL models at next-to-leading power accuracy}
\label{appendix:models}
In this Appendix we give the explicit formulas for the GW and HTL potentials up to next-to-leading power (NLP) accuracy. Using the results from Appendix~\ref{appendix:dipole}, we have
\begin{align}\label{eq:gw-expansion}
v^{\rm GW}(\x)&=\frac{\hat{q}_0}{\mu^2}\left(1-\mu |\x|K_1(\mu|\x|)\right)
\nn
&=\frac{\hat{q}_0\x^2}{4}\log\left(\frac{4e^{1-2\gamma
_E}}{\x^2\mu^2}\right)
\nn
&+\frac{\hat{q}_0\x^4\mu^2}{64}\log\left(\frac{16e^{5-4\gamma
_E}}{\x^4\mu^4}\right)+\mathcal{O}(\x^6\mu^4)\, ,
\end{align}

\begin{align}\label{eq:htl-expansion}
v^{\rm HTL}(\x)&=\frac{2\hat{q}_0}{m_D^2}\left(K_0(m_D|\x|)+\log\left(\frac{m_D|\x|}{2}\right)+\gamma_E\right)
\nn
&=\frac{\hat{q}_0\x^2}{4}\log\left(\frac{4e^{2-2\gamma_E}}{\x^2 m_D^2}\right)
\nn
&+\frac{\hat{q}_0\x^4m_D^2}{128}\log\left(\frac{16e^{6-4\gamma_E}}{\x^4 m_D^4}\right)+\mathcal{O}(\x^6m_D^4) \, .
\end{align}

Using the map given by \eqn{eq:ir-map}, the model dependent constants appearing in the generalized NLP dipole cross-section (\eqn{eq_v_NLT}), read 
\begin{equation}
c_1^{\rm GW}=64 \frac{\mu^2}{\mu_\ast^2} \, , \quad  c_2^{\rm GW}=16 \rme^{5-4\gamma_E}  \frac{\mu^4}{\mu_\ast^4}\, , 
\end{equation}
\begin{equation}
c_1^{\rm HTL}=128 \frac{m_D^2}{\mu_\ast^2} \, , \quad  c_2^{\rm HTL}=16 \rme^{6-4\gamma_E}  \frac{m_D^4}{\mu_\ast^4}\, . 
\end{equation}
\section{Details on the kinetic description of momentum broadening}
\label{appendix:kinetic_broad}
In this Appendix we outline how some of the results presented in Section \ref{sec:distribution-ioe} can be obtained solely using the kinetic description of momentum broadening introduced in Section \ref{sec:kinetic}. In addition, we provide an equivalent formulation of Moli\`ere's expansion in the form of a generalized diffusion equation in momentum space.\par 
One can rewrite \eqn{eq:kin-mom} using \eqn{eq:v} as
\begin{equation}\label{eq:ap_D_1}
\frac{\partial}{\partial L}\cP(\k,L)=-\int_\q  v(\q) \cP(\k-\q,L) \, .    
\end{equation}
In the high energy limit, single hard scattering dominates, which implies that only the first iteration of \eqn{eq:ap_D_1} contributes and thus one can write \cite{BDIM1}
\begin{equation}\label{eq:ap_D_2}
\cP^{\rm SH}(\k,L)=-\int_\q  v(\q) (2\pi)^2\delta^{(2)}(\k-\q) \, , 
\end{equation}
which is satisfied by \eqn{eq:P_SH}. On the other hand, when multiple soft scattering dominates, $v(\x)$ is quadratic in $\x$, which allows one to write \eqn{eq:ap_D_1} as a diffusion equation with diffusion parameter $\hat{q}$
\begin{equation}\label{eq:ap_D_3}
\frac{\partial}{\partial L}\cP^{\rm MS}(\k,L)=\frac{\hat{q}}{4}\vec{\nabla}^2_\k \cP^{\rm MS}(\k,L) \, , 
\end{equation}
which can be solved to yield \eqn{eq:P_MS}. \par 
Combining the kinetic description of momentum broadening with Moli\`ere's approach leads to an hierarchy of (trivially) coupled diffusion equations with a source term. To explicitly see this, we use \eqn{eq:v_Moliere} and \eqn{eq:expli_cP_series} in \eqn{eq:ap_D_1}. The leading order term satisfies \eqn{eq:ap_D_3}, while higher order terms satisfy ($\rm i\geq 1$)
\begin{align}\label{eq:ap_D_4}
\frac{\partial}{\partial L} \cP^{(i)}(\k,L)&=\frac{Q_s^2}{4L}\vec{\nabla}^2_\k \cP^{(i)}(\k,L)
\nn
&+\frac{4\pi Q_{s0}^2}{L}\int_\q \frac{1}{\q^4} \cP^{(i-1)}(\k-\q,L) \, . 
\end{align}
\eqn{eq:ap_D_4} can, in principle, be solved order by order using Green's method, but, to the best of our knowledge, and considering the aims of this paper, there is no obvious advantages over the procedure followed in the main text.

\bibliographystyle{apsrev4-1}
\bibliography{Lib.bib}

\end{document}